\documentclass{aa} 
\usepackage{graphicx}
\usepackage{txfonts}
\bibpunct{(}{)}{;}{a}{}{,} 

\def\kms {km\,s$^{-1}$}

\begin{document}

   \title{The height of convective plumes in the red supergiant $\mu$ Cep\thanks{Based on observations obtained at the T\'elescope Bernard Lyot
(TBL) at Observatoire du Pic du Midi, CNRS/INSU and Universit\'e de
Toulouse, France.}}

   \author{{ A.~L{\'o}pez Ariste}\inst{1},{ M.~Wavasseur}\inst{1}, { Ph. Mathias}\inst{1},  { A. L\`ebre}\inst{2}, { B. Tessore}\inst{3}, { S.~Georgiev}\inst{2,4}}
   \date{Received ...; accepted ...}

   \institute{IRAP, Universit\'e de Toulouse, CNRS, CNES, UPS.  14, Av. E. Belin. 31400 Toulouse, France \and
   LUPM, Universit\'e de Montpellier, CNRS, Place Eug\`ene Bataillon, 34095 Montpellier, France \and
   Universit\'e Grenoble Alpes, CNRS, IPAG, 38000 Grenoble, France \and
   Institute of Astronomy and NAO, Bulgarian Academy of Science, 1784 Sofia, Bulgaria 
}

 
  \abstract
  {}
   {We seek to understand convection in red supergiants and  the mechanisms that trigger the mass loss from cool evolved stars.}
   {Linear spectropolarimetry of the atomic lines of the spectrum of $\mu$ Cep reveals information well outside the  wavelength range expected from previous models. This is interpreted as structures in expansion that are visible in the front hemisphere and sometimes also in the back hemisphere. We model the plasma distribution together with its associated velocities through an inversion algorithm to fit the observed linear polarization.
 }
   {We find that supposing the existence of plasma beyond the limb rising high enough to be visible above it can explain the observed linear polarization signatures as well as their evolution in time. From this we are able to infer the geometric heights of the convective plumes and establish that this hot plasma rises to at least 1.1 $R_*$. }
   {$\mu$ Cep appears to be in an active phase in which plasma rises often above 1.1 $R_*$ . 
   We generalize this  result to all red supergiants in a similarly evolved stage, which at certain epochs may easily send plasma to greater heights, as $\mu$ Cep appears to be doing at present. Plasma rising to such heights can easily escape the stellar gravity. }
  

\titlerunning{Raising plumes in $\mu$ Cep}
\authorrunning{A. L\'opez Ariste et al.}

   \maketitle

\section{Introduction}

Despite their large impact on stellar and galactic evolution, the properties of outflows from red supergiants (RSGs) are not well characterized. In particular, the role of convection is still poorly understood, partly because their structures are difficult to observe directly through interferometry. In this work, we propose a view of the convection structure of the RSG $\mu$ Cep through images reconstructed from spectropolarimetric data. Convection is also of interest to those studying mass loss from these evolved stars because these convective processes are thought to contribute.



These have been the motivations for several campaigns of observation of the red supergiant $\mu$ Cep at the Telescope Bernard Lyot (TBL) with both spectropolarimeters Narval and Neo-Narval. These observations were made in parallel to those of Betelgeuse presented by  \cite{Auriere_2016}, \cite{Mathias:2018aa}, and \cite{LA18}, though
less frequently.
Nevertheless, as for Betelgeuse, the observations of $\mu$ Cep were predominantly aimed at measuring 
the linear polarization in the atomic lines of its spectrum. 

An observed net  linear polarization was interpreted as depolarization ---during
line formation--- of the spectrum continuum, which is itself polarized by Rayleigh scattering \citep{Auriere_2016}. Using this hypothesis of the physical origin of the observed linear polarization, two-dimensional images of the photosphere of Betelgeuse were inferred, images that could be favorably compared to co-temporal images of this star made using interferometric techniques \citep{LA18}. These successful comparisons suggest that the many approximations involved in this new imaging technique, which uses spectro-polarimetry, are appropriate and have spurred a further step in the technique: recently \cite{LA22} published the first  three-dimensional images of Betelgeuse.  In addition to the satisfying images, this technique provided some interesting data: the spatial and temporal scales of the convective patterns were measured, and the characteristic velocities of the raising plasma were determined. These velocities are much higher than the adiabatic estimates, easily
reaching values of 40\kms \citep{LA18,stothers_giant_2010}. More interestingly, \cite{LA22} observed that in  several observed cases of rising hot plasma, the velocity was constant with height, suggesting the presence of a force counter-acting gravity in the photospheric layers. These large velocities, sometimes reaching 60\kms, are comparable to the escape velocity at tantalizingly low heights ($1.5 R_*$). If at any time the hot plasma reaches the escape velocity, it will escape the gravity of the star and, cooling down, may be the origin of the clumpy dust clouds seen around Betelgeuse \citep{montarges_dimming_2021}. This fast, rising plasma may also be the source of mass loss in these stars. 

However, in order to reach this interesting result, a critical piece of information is missing. The technique presented and used by \cite{LA22} to build three-dimensional images of the photosphere of Betelgeuse is unable to determine the geometric height of the successive layers imaged. The technique only provides the ordering of the layers, from the deep atmosphere up to higher layers, but not the geometric distance between them. Here, we present a means to measure or at least estimate this geometric height.

In Section 2 we present the set of observations of $\mu$ Cep collected  with Narval and Neo-Narval at the TBL from 2015 through 2022.
 In Section 3, we describe the spectral features in the linear polarization of $\mu$ Cep that cannot be explained with the model used to image Betelgeuse, and propose a modification of the model. We propose that, from time to time, convective plumes are powerful enough to rise to sufficient heights that they can be seen beyond the geometric horizon of the star.  This cannot be a permanent feature, but it may happen from time to time, an aspect that is critical to the   plausibility of this proposition. We also discuss how this modification affects previous results for Betelgeuse, if we assume that a unique model serves all RSGs.
In Section 4 we build an inversion code based on this modified model, where the usual description of the brightness variation across the disk is supplemented with the presence of up to five clouds of plasma visible beyond the horizon of the star. The measured linear polarization degrees and angles allow us to determine how far beyond the horizon this plasma is and therefore how high it must be to be visible from Earth. It is in this way that we can determine a minimum height for these structures. In Section 5, we build a time evolution of one of those plumes that we were lucky to follow  in 2021 from rise to fall. In Section 6, we put these measurements in context, in particular with  respect to the measured plasma velocities. We confirm that it is highly possible that the most powerful of these convective plumes are high enough to escape the gravity of the star at the observed velocities.

\section{Spectropolarimetric data from  Narval and Neo-Narval}

$\mu$ Cep is an M2-type RSG with stellar parameters ($T_{eff} = 3750 K$ and $\log g = -0.36$ ) very similar to those of Betelgeuse, while its mass (25$M_\sun$)  and radius (1420 $R_\sun$) on the other hand may be larger \citep{levesque_physical_2005}. \cite{tessore_measuring_2017}  first detected strong linear polarization features (both in Stokes $Q$ and  $U$) associated to atomic lines.

We began observing  $\mu$ Cep in linear polarimetry in July 2017 with the Telescope Bernard Lyot at Pic du Midi (France,TBL). Until August 2019, the Narval spectropolarimeter was used. After an upgrade, starting in September 2019,  Neo-Narval re-observed $\mu$ Cep in May 2020 and regular observations have been conducted ever since. This long series allowed us to follow the entire lifetime of one of these convective plumes (see Sect. 5).

Narval and Neo-Narval have been described before in the literature. These descriptions are extensive for the case of Narval \citep{Donati2006}, with a succinct description of the changes  of Neo-Narval provided by \cite{LA22}.  As stressed in this latter publication, we note the continuity of data quality from the instrument through its upgrade, and handle Narval and Neo-Narval data as a unique dataset with no further reference to the instrument used.

We performed relatively short exposures (about 3 minutes per polarimetric sequence) in order to ensure a peak signal-to-noise ratio (S/N) of about 2000 in Stokes I per velocity bin.
A list of  the observations of $\mu$ Cep  is presented in Table \ref{tab1}, corresponding to all those studied in this work. A least-squares deconvolution (LSD) procedure \citep{Donati2006} is applied to the reduced spectra. Atomic lines from an appropriate list \citep{Auriere_2016} are summed after rescaling of the wavelength binning. The result is a single spectral profile for both Stokes I and the observed Stokes parameter.
The whole set of Stokes Q, U,  and I profiles thus obtained is  shown in Fig.\ref{velos} in the form of an image with time on the vertical axis.

\section{Redshifted linear polarization features.}
The model used to interpret the linear polarization observed in the atomic lines of Betelgeuse and $\mu$ Cep and, in general, of all red supergiants assumes a nonrotating convective star. This model and the implicit approximations involved were described in detail by \cite{LA18} and \cite{LA22}. From the point of view of the physical origin of the linear polarization, our model assumes that what we observe is the depolarization of the continuum  by atomic lines due to Rayleigh scattering. A key diagnostic of the trustworthiness of this interpretation of the polarization is that all lines must show similar polarization, independent of their quantum structure. In particular, the Na $D_1$ and $D_2$ lines must show similar signals to one another. This was seen to be the case for Betelgeuse \citep{Auriere_2016}, CE Tau, and now for $\mu$ Cep,  the target of the present study. Once the physical  origin of this polarization is confirmed,  the model focuses on the distinct spatial origin of the spectral features seen in the linear polarization profiles.

The observed linear polarization profiles characteristically show several distinctive lobes inside every atomic line. In the absence of rotation, the wavelength position of each of those lobes is assumed to be due to convection. The brightest plasma is assumed to be rising, and the cooler, darker plasma sinks. At first approximation, most light comes from the brighter regions and is therefore Doppler shifted by the projection onto the line of sight of the convective velocity at which the plasma rises. Thus, bright hot plasma at disk center will emit light in the blue wing, while bright hot plasma at the limb will emit light in the red wing, at a wavelength which will coincide with the velocity of the center of mass of the star with respect to the Sun. Dark, sinking plasma would be redshifted with respect to this red wing, but its low intensity translates into a  tiny signal  to be added in the further red wing of the observed profile. In this model, the spectral profile of an atomic line is framed by two velocities. One of these velocities is the heliocentric velocity $V_*$, which limits the red wing of the polarization profile. 
Plasma at the stellar limb will emit light at or near to this red wavelength. The second of these velocities is the maximum velocity of the plasma in the convective flow, $V_p$, which limits the blue wing of the profile. Plasma rising at this maximum velocity  at disk center will emit light at the bluest wavelengths.  In the absence of rotation, all other velocity fields, such as micro- and macroturbulent velocities or thermal broadening,  are assumed to be isotropic and would just broaden the signals. Such broadening, added to instrumental effects, is seen as a minimum width for all observed polarization signals, a width that is much smaller than the span of velocities attributed to convection.

\begin{figure}
\includegraphics[width=0.5\textwidth]{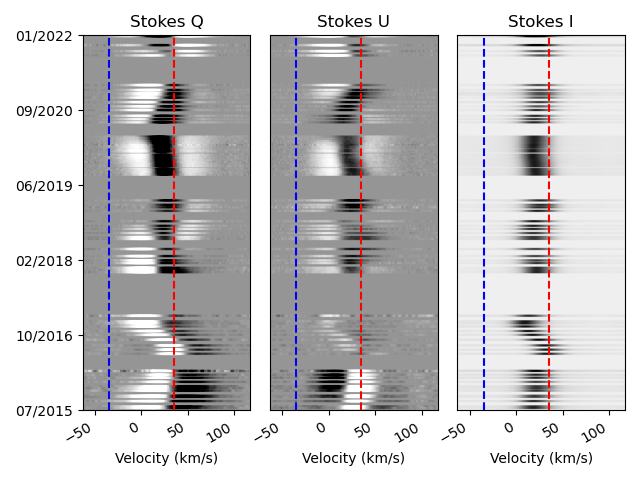}
\caption{Pile-up of the Stokes Q (left), U (center) and I (right) profiles over the whole time series. For illustrative purposes, every observation has been made to span 15 days on the vertical direction. The blue and red vertical lines mark the maximum plasma velocity $V_p$ and the radial velocity of the center of mass of the star $V_*$ respectively (see main text for definitions). Velocities are measured in the heliocentric reference system. }
\label{velos}
\end{figure}

The two velocities, $V_*$ and $V_p$, that limit the observed profiles are parameters of the model and should be determined a priori. As discussed by \cite{LA22}, this a priori determination is done by inspection of the whole set of available observations. In Fig. \ref{velos}, the two velocities are represented as vertical lines on top of the pile up of the observed profiles. Our choice for these two velocities, namely $V_*=+35$ \kms for the velocity of the center of mass of the star and $V_p=-70$ \kms for the maximum velocity of the convecting plasma, can be judged with respect to the wavelength span of the polarization signals. We note that while $V_*$ is measured in the heliocentric reference system, we are giving the value of $V_p$ in the star's own reference system. In the heliocentric reference system used in Fig. \ref{velos}, we find $V_p$ at $+35-70=-35 $\kms. As $V_p$ has a meaning in terms of the physics of convection of the star, it is useful to keep its value in the reference system of the star, even at the risk of some confusion when looking at Fig.\ref{velos}.

 The choice of these values is not free from criticism. Judging from Fig.\ref{velos} alone, it appears as if the red limit $V_*$ has been placed in the middle of the polarization signal rather than at its red edge.  As these two velocity limits cannot be directly measured, we can only advance the arguments that justify our choice for these two parameters.   These arguments are qualitatively similar to the ones used by \cite{LA18} and that \cite{LA22}
 justified to be acceptable within 10 \kms. Part of this justification lies in the fact that accepting the model and the values of these velocities results in images of Betelgeuse or CE Tau, another observed RSG, that are comparable with contemporaneous images inferred by interferometers \citep{LA18}.  However, in the present case, we lack any such interferometric images for $\mu$ Cep, and  contrary to the previously studied Betelgeuse and CE Tau, there is a considerable amount of signal redward from $V_*$. It is worth examining the arguments that justify this choice.

 It is obvious that $V_p$, the maximum velocity of the plasma represented by the blue line  in Fig.\ref{velos} (which, we reiterate, is in the heliocentric reference system, while the value of $V_p$ is given in the star's reference system), must encompass the most blueshifted signals observed over the years. Accordingly, added to $V_*$, this velocity must be somewhere beyond -20\kms in Fig.\ref{velos}.  We have chosen -35\kms to include the extended wings of the signals observed.  In our model, we have no explanation whatsoever for any signal blueward from $V_p$. We therefore have to make sure that there is no signal beyond this limit, and this fixes minimum values of $V_p$.

In our model, $V_p$ is interpreted as the velocity of the rising plasma during convection in the reference frame of the star. This interpretation sets further constraints on its maximum value. Our choice has been, for $\mu$ Cep, to set this maximum velocity at $V_p=-70$ \kms. This is already a large velocity for rising plasma and is roughly seven times the speed of sound in the atmosphere of $\mu$ Cep. Numerical simulations \citep{freytag_spots_2002, chiavassa_radiative_2011} produce supersonic flows  for convective patterns,  which were confirmed by \cite{LA18}. But  a value of  $V_p=-70$ \kms is 
50\% to 100\% larger than any published figure so far, based on either observations or numerical simulations.

However, as $V_p$ is fixed on its blue side by the extent of the polarization signal, accepting these large plasma velocities is the only way to place $V_*$  as far to the red as possible.
As $V_*+V_p$ is set at -35 \kms,  the velocity of the center of mass must therefore be  $V_*=+35$\kms.  
In spite of the large value of $V_*$, this choice  still leaves lots of redshifted signal beyond the limit of the model. Once again, including those signals in our convection model by shifting $V_*$ to higher values would imply accepting convection velocities larger than 70\kms and this seems unphysical. Also, once more, diminishing the maximum velocity $V_p$  also seems unphysical, as blueshifted signatures would be left unexplained beyond the maximum velocity of our model. This sequence of arguments justifies our choices up to 10\kms , and leaves large amounts of signal beyond the red limit of $V_*$.
 
 Another argument justifying the choice of  $V_*=35$\kms is apparent from  the Stokes $I$ variation.
From the LSD profile, one can compute a mean heliocentric radial velocity from the profile Gaussian fit. This mean velocity is found to be  $<v>=22$\kms \  in the heliocentric reference frame, and from profile to profile varies with an amplitude of about 4\,km\,s$^{-1}$.
For Betelgeuse, the corresponding quantities are respectively $<v>=21$\kms and 4\kms .
The radial velocity of Betelgeuse was estimated to be about $V_*=40$\kms.
Supposing the same parameters for both these RSGs, that is, a number of granules number of similar order and the same temperature contrast, then the
$V_* - <v>$ values should also be similar, meaning a value of the order of 40\kms  , or, within
the 10\kms uncertainty, the adopted $V_*=35$\kms value.
To conclude this discussion of the choice of the value of these two velocities, we must add that several velocities have been tested inside the range allowed by those 10 \kms, without significant differences in the results. 

A strict interpretation of these  velocity limits  implies that no polarization signal can be seen  in our modeled profiles at wavelengths redder than $V_*$, the limit given by the velocity of the center-of-mass of the star for each atomic spectral line. Polarization signals beyond this limit would come, in this model, from plasma moving away from the observer and towards the center of the star. Plasma sinking towards the core of the star is assumed to be cool, dark plasma. This rigorous interpretation must be softened somehow, as cold, dark plasma still emits some light and plasma may start sinking while still being bright enough to contribute to the net spectral line profile; but these are always small contributions, and we are not expecting any large signals on the red side of the velocity limit. After inspection of the profiles of Betelgeuse published by \cite{Auriere_2016}, \cite{Mathias:2018aa}, \cite{LA18} and \cite{LA22} we can confirm that this is the case, and that the hypothesized model can confidently describe all the available observations.  However, this is not the case for $\mu$ Cep. 

\begin{figure*}
\includegraphics[width=\textwidth]{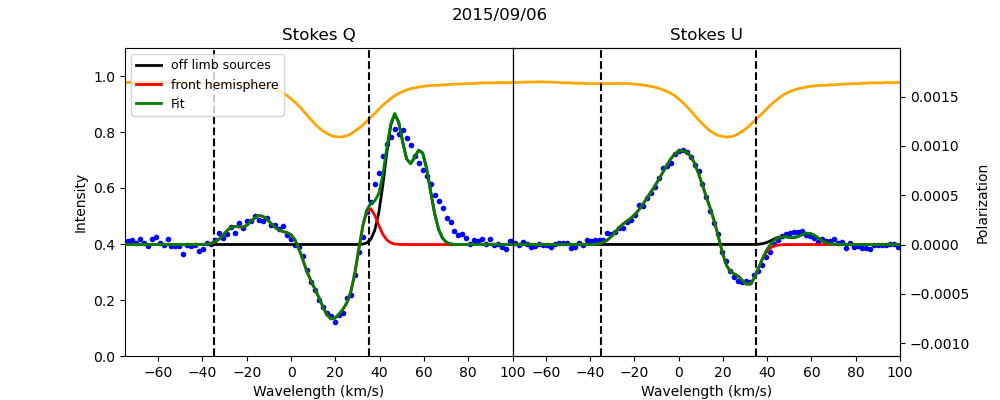}
\caption{Observed linear polarization of $\mu$ Cep on September 6, 2015. The observed Stokes Q is plotted on the left, and Stokes U on the right,  as
dots in both cases. Continuous lines represent the best fit from the assumed model (green line) with separated contributions from the front disk brightness distribution (red) and the two plumes beyond the limb (black), visible at those wavelengths when they do not constitute the whole contribution to the final fit in green. The upper (orange) profile shows the normalized intensity profile. The vertical dashed lines give the two limiting velocities, $V_p$ and $V_*$.}
\label{prof1}
\end{figure*}

In  Fig. \ref{prof1}, we plot the observed spectra of $\mu$ Cep collected since September 6, 2015. The observations are plotted as dotted lines. We return to the continuous lines in different colors further below. At this point, we focus our attention on the strong Q signal peaking at about +50 \kms. This is actually the strongest polarization of the observed profiles on that date and it is found to the red of the limiting velocity  $V_* =+35$ \kms, and is indicated by the vertical dashed line.

Such strong signal in the red wing of the profiles of atomic lines is unlike anything observed  to this day in Betelgeuse. The many more observations of  linear polarization in the spectra of Betelgeuse are better illustrated by the profile of Stokes U shown in  the right plot of Fig.\ref{prof1}. A strong peak is seen in the blue wing and is attributed to bright plasma near the center of the stellar disk; another (negative) peak is seen near $V_*$, and attributed to bright  rising plasma coming from regions near the stellar limb; and a small   (positive) signal is seen beyond $V_*$ corresponding, hypothetically,  to sinking dark plasma. The observed Stokes U profile is, in this manner, qualitatively explained and, after inversion, the inferred image confirms this basic description of the visible structures. Such a model would also explain the small lobe seen in the blue wing of the Stokes Q profile and the larger negative peak near the red boundary (red line of Fig. \ref{prof1}). The respective amplitudes and signs of these peaks in Q and U will constrain the position and brightness of the different bright structures over the disk. However, this model has no explanation whatsoever for the strong signal on the red side of the red boundary of the Q profile. Such strong signal cannot be attributed to dark sinking plasma, for there would be no explanation for its large amplitude. The amplitude could either be due to the amount of photons or the polarization degree of those photons. To interpret such a large amplitude would require either a brighter region or a more polarized region. It appears contradictory to say that the sinking plasma is brighter than the rising plasma, and so we are only left with the possibility that this is sinking plasma with an enormous polarization degree. Implicit in our model is that polarization degree is directly related to the height of the plasma. One possible explanation that our model would have for this strong polarization peak would therefore be that this is a huge cloud of cold plasma sinking from large heights that is much larger than any other structure in the atmosphere, because height must compensate for the loss of signal due to the lower emissivity of this cool dark plasma.

The presence of that unexpected strong peak is forcing the model towards extreme scenarios.  

Another possibility is that our determination of $V_*$, the velocity of the center of mass of the star, is wrong. It suffices to shift this limit a further 35 \kms to the red, up to $V_*=+70$\kms, for the entire peak to fit inside the limits of the model. However, this scenario  entails unpleasant conclusions as well. Shifting this limit to the red  without touching the blue limit would mean that the maximum velocity of the convective flows in $\mu$ Cep would increase to a staggering 100 \kms. This is an uncomfortably large number for the convective flows, about ten times the speed of sound. The problems with a modified red boundary do not end here. We expect there to always be some signal coming from near the limb, because statistically there is a large probability of finding a bright structure somewhere along the long circumference. This is the case with the actual red boundary limit plotted in Fig. \ref{prof1}: both Stokes Q and U profiles show signal near the limb. However, if we were to accept this boundary shift, Stokes Q would still have the strong peak that would be attributed to a near-limb structure, but no comparable signal is visible anywhere near that limit for Stokes U. In order to produce such signal imbalance between Q and U, one would need to imagine a stellar disk with a continuous dark band along and inside the limb except at one position where a bright structure would give the observed Stokes Q signal. While not impossible, this appears as a strange disposition of structures on $\mu$ Cep, something never seen on Betelgeuse. In addition, this 70 \kms value is clearly outside the I profile, meaning that this latter would have no link with the heliocentric star velocity, which also seems difficult to accept.

\begin{figure*}
\includegraphics[width=\textwidth]{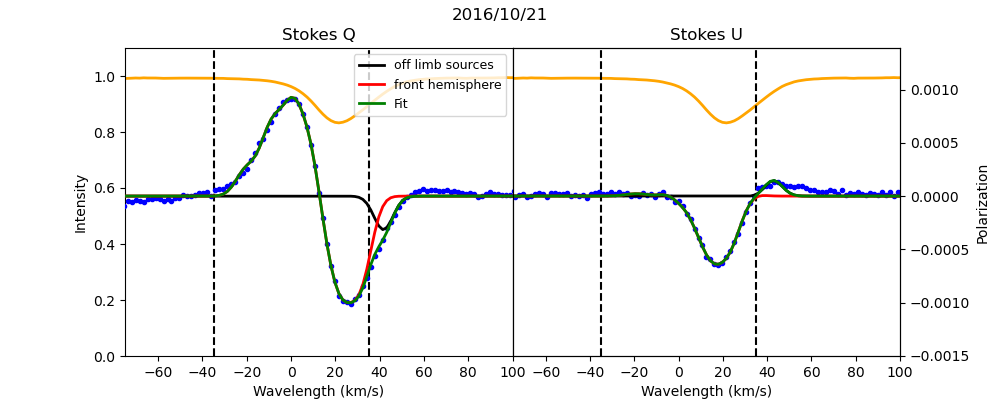}
\caption{Observed linear polarization of $\mu$ Cep on October 21, 2016. The same color codes are used  for Stokes Q and U as in Fig. \ref{prof1}. The velocity boundaries given by the dashed lines are common to all the observations of $\mu$ Cep.}
\label{prof2}
\end{figure*}

One year later, in October 2016, the observations shown in Fig.\ref{prof2} have drastically changed. Between the velocity boundaries, the polarization signals keep providing an image of changing bright, convective structures, but there is always signal at the qualitatively expected places, even if that signal has changed in amplitude, ratio, and position. This is interpreted as bright structures that have moved over the disk; some have appeared anew, others disappeared, but there is always signal coming from around disk center and visible around the blue wing, and signal coming from the limbs and visible around the red wing, but on the correct side of the boundary $V_*$. The big change is that at this date, and contrary to the observations in 2015, there are no conspicuous signals on the wrong, red side of $V_*$. There are always small amplitude signals, both in $Q$ and $U$. Because of their small amplitude, they can be comfortably assigned to dark sinking plasma, or perhaps a small error in the determination of $V_*$, an error of at most 10 \kms, which is consistent with the rough arguments used in its determination. However, there is no large peak visible, which casts doubt on the validity of the model.

These two observations of $\mu$ Cep show an expected signal between the velocity boundaries that, while changing, is always there. It can be explained as it was for Betelgeuse: by a spatially inhomogeneous distribution of bright patterns that have been interpreted as convection. However, these observations also show a new signal  that appears and disappears in time, and that, if we accept the velocity boundaries, corresponds to bright plasma moving away from the observer.

The conclusions drawn from  these qualitative arguments are definitively confirmed by the inversion codes developed by \cite{LA18} and \cite{LA22}: the model used to fit the observed polarized spectra of Betelgeuse 
and to infer the published images is unable to produce a solution for the spectrum of $\mu$ Cep on September 2015, though it provides a solution for the observations of October 2016. Unwilling to discard a model that has been successful with Betelgeuse, we propose an addition to this model that can explain the intermittent appearance of strong signals on the red side of the red velocity boundary, which are illustrated in Fig. \ref{prof1}. We propose that the bright convective structures inferred for Betelgeuse and $\mu$ Cep and present over the whole star ---and also in the back hemisphere---,  may raise plasma high enough for it to become visible above the stellar limb. We refer to this high rising hot plasma  as plumes.
When these plumes are on the front hemisphere, they produce the signals between the two velocity boundaries and the basic model is able to explain them. Similar convective bright structures must also occur on the back hemisphere, but they are usually hidden by the stellar limb. From time to time, one of these bright structures in the back hemisphere may push plasma high enough for it to become visible to us above the limb. This plasma rises in a radial direction, but as it is in the back hemisphere of the star, we see it redshifted, moving away from us beyond the red velocity boundary; it is bright plasma nevertheless, and so we expect it to have similar polarization amplitudes to plasma in the front hemisphere in symmetric geometries. Plasma is not usually expected to rise high enough, and so we often expect to see nothing beyond $V_*$. This has been the case for all available observations of Betelgeuse and also for $\mu$ Cep in October 2016. However, from time to time this may happen, producing the signal illustrated in Fig. \ref{prof1}.  When this is observed, it cannot happen all over the stellar limb, but only at particular polar angles, thus explaining the single peak that is visible only in Stokes $Q$.    

Becoming visible over the limb depends  on the distance to that limb of the bright structure.  The further a structure is from the limb, the higher it has to rise to become visible.  This suggests that we can determine the height of one of those structures as the minimum height at which, by geometry,  they become visible above the limb.  This measurement of a minimum height for the rising plasma to become visible is going to be our main result.

\section{Inversions with a modified model}

\begin{figure}
\includegraphics[width=0.5\textwidth]{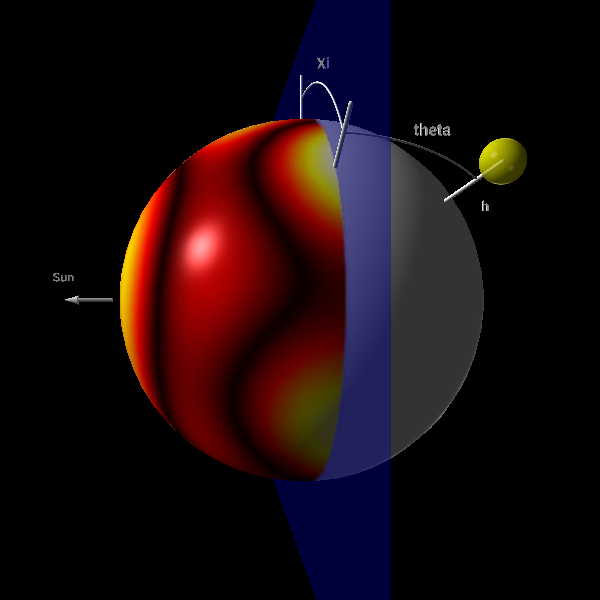}
\caption{Cartoon defining the parameters of a discrete source (yellow sphere) in the back hemisphere (in gray) beyond the plane of the sky (bluish plane). Celestial north is up, in the plane of the sky. The image of the front hemisphere corresponds to the inferred brightness distribution of $\mu$ Cep in September, 2015.}
\label{cartoon}
\end{figure}

In accordance with the suggested modification of the model proposed in the previous section upon inspection of those polarization signals beyond the red velocity boundary, we built an inversion code to fit the observed spectra of $\mu$ Cep. The core of this inversion code is identical to the one described by \cite{LA18}. Mathematically, it is a Marquardt-Levemberg algorithm that fits the observed Stokes Q and U profiles with synthetic profiles computed from a distribution of brightness over the surface of  the star. On the front hemisphere, this distribution of brightness is described by a linear combination of spherical harmonics up to sixth order. The blue velocity boundary is the maximum velocity of the rising plasma. The brightest point over the disk at any particular realization of the model is supposed to move radially at that maximum velocity. All other points over the disk have a brightness described by the spherical harmonics, and a velocity which is mathematically related to its brightness, meaning that the resulting brightness contrast and velocities roughly match the solar case \citep[see the Appendix in][]{LA22}. The polarization emitted by a point over the hemisphere is proportional to that brightness, but also to the squared sine of the scattering angle, as expected for Rayleigh scattering. The ratio of polarizations between Q and U is given by the tangent of half the polar angle position of the point. Its wavelength is determined by its velocity projected onto the line of sight and therefore depends on the distance to the center of the disk. Mathematically, the model uses, as parameters,  the coefficients of  a brightness distribution written in terms of spherical harmonics as
\begin{equation}
B(\mu,\chi)=\left \| \sum_{\substack{\ell=0,\ell_{max} \\ m=-\ell,+\ell}} a_{\ell}^m y_{\ell}^m(\mu,\chi) \right \|
\label{Beq}
,\end{equation}
with $\ell_{max}=6$, and $\mu$ and $\chi$ being the angle to disk center and the polar angle with respect to celestial north, respectively. This brightness distribution results in the emission of net polarization described in terms of Stokes parameters as
\begin{eqnarray}
Q_{disk}(v)=\sum_{\mu,\chi,v_z} B(\mu,\chi) \sin^2 \mu \cos 2\chi e^{-(v-v_z)^2/\sigma^2},\\
U_{disk}(v)=\sum_{\mu,\chi,v_z} B(\mu,\chi) \sin^2 \mu \sin 2\chi e^{-(v-v_z)^2/\sigma^2}
\label{QU}
,\end{eqnarray}
where $v_z=V(\mu, \chi) \cos \mu$, with $V(\mu, \chi)$ being the plasma velocity at that point, and proportional to the brightness and  limited by the maximum speed of the plasma $V_p$. Each emission is broadened with a Gaussian profile of fixed width $\sigma=6$ \kms  (i.e., 10 \kms FWHM), which represents both instrumental and thermal broadenings.
This is a rough description of the basic model, for which many more details are given and scrutinized by \cite{LA18} and \cite{LA22}.
In addition to  this basic model, we assume the presence of one or several sources of polarization beyond the limb. When adding new parameters to the model to describe those new sources of polarization, one should be careful not to overload the inversion algorithm with more new unknowns than available new information. Therefore, it would be unreasonable to try to provide a description of the continuous distribution of brightness in the back hemisphere, because only a very limited amount of that plasma will be contributing to the observed spectra. It is tempting to try to propose a description of the brightness in a ring above the limb.  Unfortunately, we have not found a proper mathematical description for such a ring. One of the difficulties is that, as we assume that it is uncommon for plumes to  rise high enough to become visible above the limb, we are expecting contributions from at most a small range of polar angles, the rest contributing zero. Any orthogonal family of functions trying to describe this paucity of sources requires an excessive number of parameters. We finally opted for a simplistic description in terms of a small number of discrete sources. Each one of these discrete sources over the limb is described by its polar angle $\chi$, its angular distance to the limb $\theta,$ and a brightness value $Z$ (see cartoon in Fig.\ref{cartoon}). Its polarization is given, as in the case of any other emitting point in the front hemisphere, by the scaled product of its brightness and the squared sine of the scattering angle, this scattering angle being geometrically related to the distance to the limb.
\begin{eqnarray}
Q_{\mathit{off}}(v)=\sum_{i=0,N} Z_i \sin^2 \theta_i \cos 2\chi_i e^{-(v-v_z)^2/\sigma^2},\\
U_{\mathit{off}}(v)= \sum_{i=0,N} Z_i \sin^2 \theta_i \sin 2\chi_i e^{-(v-v_z)^2/\sigma^2}
\label{QU_off}
.\end{eqnarray}

The radial velocity of the rising plasma is identically given as a function of brightness.  Its redshifted wavelength is analogously given by the projection of this velocity onto the line of sight, a projection which is again geometrically dependent on the distance to the limb. Each one of these discrete over-the-limb sources produces a polarization peak in Stokes Q and U that is broadened by a Gaussian profile with a full width at half maximum (FWHM) of 10 \kms. This FWHM is supposed to encompass both the instrumental resolution and various stellar broadening mechanisms, that is, thermal, microturbulent, and so on.  Finally, both sources of polarization, $Q_{disk}, U_{disk}$, and $Q_{\mathit{off}}, U_{\mathit{off}}$ are added.

The last parameter to be determined is the number $N$ of discrete sources to be allowed. We find it unpractical to leave this number unbound, and prefer to fix it. We attempted from $N=0$ up to $N=5$ discrete sources. As expected, having zero sources allows us to recover the basic model, unable to reproduce the anomalous signals on the red wing. On the other end, we find that beyond four sources we are not learning anything new from the inversion results, and the algorithm becomes unstable, and presents convergence issues. This can be safely understood as an excessive number of new parameters given the available information.   Between one and four sources is therefore the right number of sources that we can safely infer. Interestingly, we also find that for  any individual  observation of our long dataset, the inferred value of the polar angle of all the sources is similar, even if the intensity and height of each one of them is different. This means that the solution found by the inversion algorithm proposes that, at a given polar angle, there are several sources of polarization at different heights and with different intensities. That is, the inferred sources clump together on the same region above the limb. This can be interpreted as one single but extended source  over the limb of $\mu$ Cep at the time of the observation. This result appears to justify our intuition that such events of high-rising plasma are not common.  From this conclusion, one may expect one single source in our model to be sufficient to describe the observed polarization profiles in the red wing, but we find that this is not the case and that we need a minimum of two sources to reproduce the basic spectral features observed. This may be an indication that even if there is a unique object beyond the limb, it has sufficient structure that our description in terms of a Gaussian profile per source is inadequate.  Using two or more sources becomes a simple manner of better describing the extent and structure of the emitting region. Because of this result, we present inversions in this work with just two sources. This has the advantage of capturing the important physical parameter for our work, the main distance of the bright structure to the limb,  while easing the constraints on the inversion algorithm. The observed structure is  often spectrally broader than twice the FWHM of 10 \kms of every discrete source. The fit is therefore approximative. By increasing the number of sources, we improve this fit, but do not bring any further information.   

The above developments are illustrated in Figs. \ref{prof1} and \ref{prof2}. Both figures show on top of the observed profiles the solution found by the inversion code as a green continuous line. This solution is made of three different contributions. The basic model describing the front hemisphere as a linear combination of spherical harmonics is plotted in red, and is fully coincident with, and hidden behind,  the full solution between the two dashed lines that limit the contribution of the front hemisphere. This basic model can only be seen as a tail of small signal on the red side of the red velocity boundary. This small signal, as mentioned above, is the contribution from the dark sinking plasma, and is insufficient to explain the observed polarization peak in September 2015. However, it is almost sufficient to explain the entirety of  the redshifted signal in October 2016. The two other contributions combined are shown as a black continuous curve, and correspond to two discrete sources above the limb. Again, this black line is only visible when it does not fully coincide with the final solution plotted in green.
As explained, limiting the number of sources to just two results in an approximative fit of the redshifted signal. The full solution profile clearly shows two peaks on the red wing, coinciding with the maxima of the two sources, a feature absent in the observations. There is also a clear tail further towards the red in the observations that cannot be captured with just two sources. Adding more sources would correct these missed fits, but the parameters of the added sources will not change significantly. In September 2015, the two sources over the limb bring signal comparable to anything else over the front disk. In October 2016, the two sources appear as small contributions that may drop to zero if just the red velocity boundary is shifted a few \kms \ towards the red. The modified model is therefore able to capture both those cases with important sources over the limb as well as those cases with negligible contributions.

We used this model, in conjunction with two sources above the limb, to invert the whole available dataset of linearly polarized spectra of $\mu$ Cep presented in Sect. 2. Imaging from linear spectropolarimetry is subject to a certain number of ambiguities: Several images, with different distributions of brightness are compatible with the same observed polarized spectra,
that is, they are possible alternative solutions of the inversion problem. These latter images are not completely unrelated. The most common ambiguity appears between two images that are identical but rotated 180 degrees with respect to one another. A comparison with images of Betelgeuse made with interferometric techniques allows us to determine which of these two rotated images is the one that better corresponds to reality. However, we do not have interferometric images for all dates, and none for $\mu$ Cep. Because of this, in the case of Betelgeuse, the best solution for a date with available interferometric images is propagated as the initial solution to the next date, which encourages the inversion code to stay in the group of solutions sharing choices among the possible ambiguities that better compared to interferometric images at one particular date. Similarly, for $\mu$ Cep, we inverted the first available date without constraints. But for next dates, the solution of the previous available date was used as initial condition. This ensures a certain time coherence in the series of images. 

\begin{figure*}
\includegraphics[width=\textwidth]{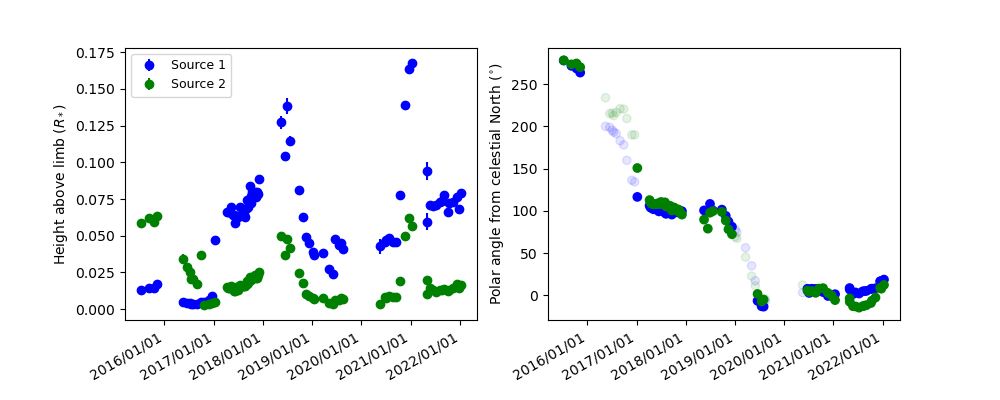}
\caption{Plots of  the height of the two sources visible above the limb and of their polar angle for the observations of  $\mu$ Cep over the last six years. For heights below 0.05, the source is considered to be absent and the corresponding value of the polar angle is made transparent.}
\label{height}
\end{figure*}

The inversion code provides  values for the polar angle and distance to the limb for the two sources over time. The distance to the limb $\theta$ is directly converted into the minimum height $h$ above the stellar surface for this source  in the back hemisphere to be visible above the limb: 
\begin{equation}
h=\frac{1-\cos \theta}{\cos \theta} R_{*}
.\end{equation}
Presented in this manner, the results of our inversions are shown in Fig. \ref{height}

Over the last six years, $\mu$ Cep appears to have produced three events in the back hemisphere with plasma being lifted to considerable heights. The first of these events was ongoing when our observations started in September 2015. It had completely disappeared when the star was re-observed in late spring 2016. The next event started one year later, between January and April 2017, and by January 2018 plasma had reached heights of  at least 1.1$R_*$ and perhaps higher. This plasma appeared at polar angles of 100 degrees and after the winter blind window, the plasma was still at the same position and at even greater heights of 1.15$R_*$. Over the spring of 2019, the emitting plasma was seen to be descending in height, until it disappeared by the summer of that year.  The beginning of the observations with Neo-Narval  at the beginning of  2020 showed $\mu$Cep to be still quiet, with no particular signals on the red wing. But this situation changed  by the end of the year, with the rapid rise of a new clump of plasma at a polar angle of 0 degrees ---and therefore unrelated to the previous one---, which in less than one month reached heights of at least 1.175$R_*$. The maximum height reached by this event appears to be quite ephemeral. As fast as it rises, it disappears. But as it disappears, we are left with a low-lying clump that persists throughout 2022. Optimistically, we may interpret this as a large event of rising plasma inside of which there is a small clump at high speed reaching even higher heights in a short time before disappearing, perhaps due to a quick cooling, while the rest of the rising plasma is still visible.  In all these events, the value of the polar angle of the two sources is quite similar, as can be seen in the right plot of Fig. \ref{height}. As said above, we interpret this result as proof that there is a unique source above the limb but more extended and complex than what our model with two Gaussians can reproduce. 

Our 8 years of observations of linear polarization of Betelgeuse have not produced any single event sufficiently large to require a modification of the inversion model. In 5 years of observations, $\mu$ Cep  has produced three such events. It is possible that this is due to the slightly different stellar parameters of these two stars. The fundamental parameters of $\mu$ Cep recently determined by \cite{montarges_noema_2019} show a star similar to Betelgeuse within  error bars. Rather than invoking fundamental differences between the two stars, we speculate that $\mu$ Cep may be at present in a \textit{Decin stage} \citep{decin_probing_2006}, as suggested by  \cite{montarges_noema_2019}, with common episodes of mass loss, while Betelgeuse may rather be in a quiet stage with rare and separated events of this kind. This is simply speculation. At this point, we lack any clear scenario explaining why and when a red supergiant will enter into a \textit{Decin stage}, if such episodes exist at all. Further observations in time will be needed  to see whether or not $\mu$ Cep stops producing these events\footnote{It must be said that    \cite{decin_probing_2006} estimate the duration of such episodes in the tens of years.}. 

In the fall of 2019, Betelgeuse suffered a large dimming that has been attributed to the formation of a dense dust cloud almost directly along our line of sight \citep{montarges_dimming_2021}. \cite{LA22} suggested that these mass-loss events are triggered from fast-rising plasma in the photosphere reaching the escape velocity at a certain height. The suggestion made by these latter authors stems from the measurement of plasma velocities that are constant with height and sufficiently large to be comparable to escape velocities at the estimated heights of these structures. It is tempting to see this event in Betelgeuse as one example of the more common events in $\mu$ Cep of plasma rising sufficiently high to be visible above the limb. But in the case of Betelgeuse, this event happened in the front hemisphere, rather than in the back hemisphere as in $\mu$ Cep. If we accept that Betelgeuse is at present in a quiet stage of mass loss, unlike $\mu$ Cep, events where plasma is ejected from the star appear to still happen.
Just  by chance,  in Betelgeuse, lately, they have not been happening in the regions near the limb, but rather in the front disk, the ultimate example being the one that produced the large dust cloud involved in the great dimming of 2019. In $\mu$ Cep on the other hand, three such events have taken place in regions around the limb, making it visible to our spectropolarimetric measurements.

\section{Follow-up of a convective plume above the limb}

\begin{figure*}
\includegraphics[width=0.5\textwidth]{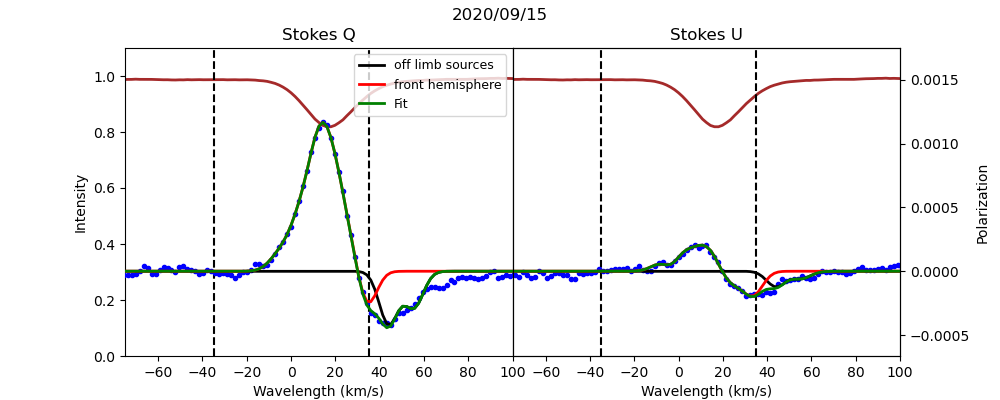}
\includegraphics[width=0.5\textwidth]{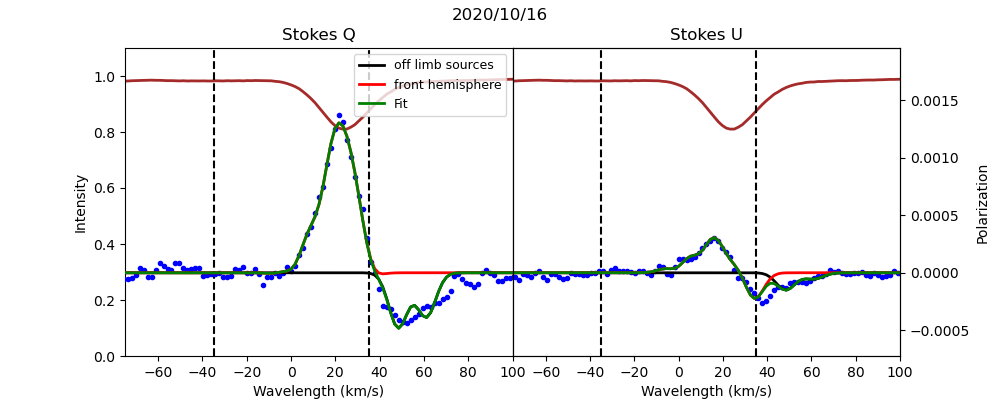}
\includegraphics[width=0.5\textwidth]{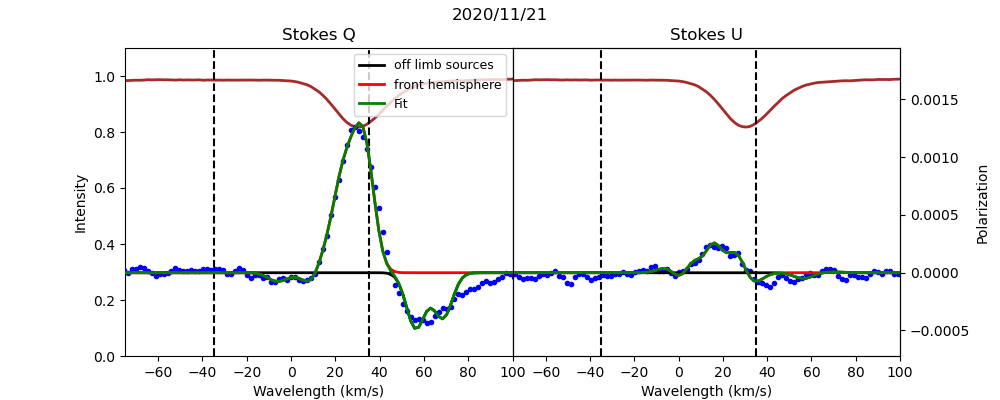}
\includegraphics[width=0.5\textwidth]{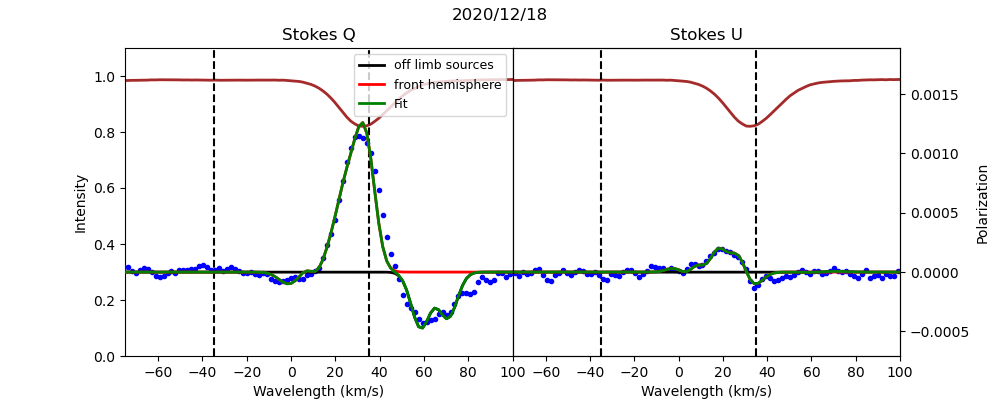}
\includegraphics[width=0.5\textwidth]{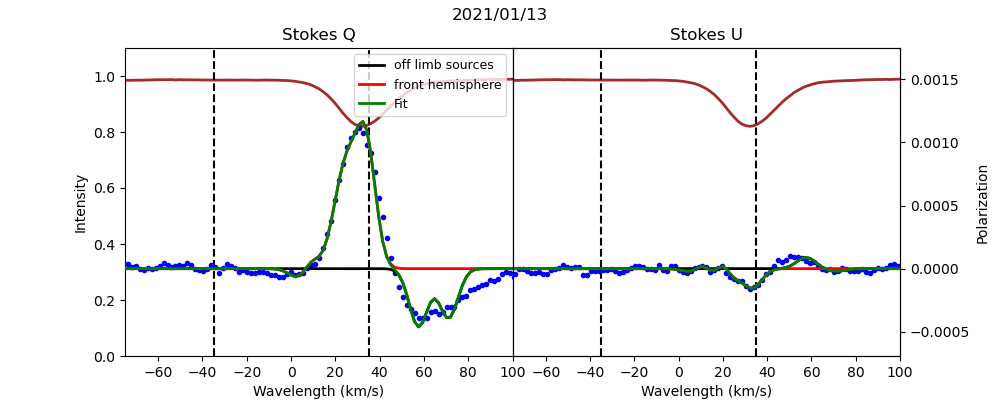}
\includegraphics[width=0.5\textwidth]{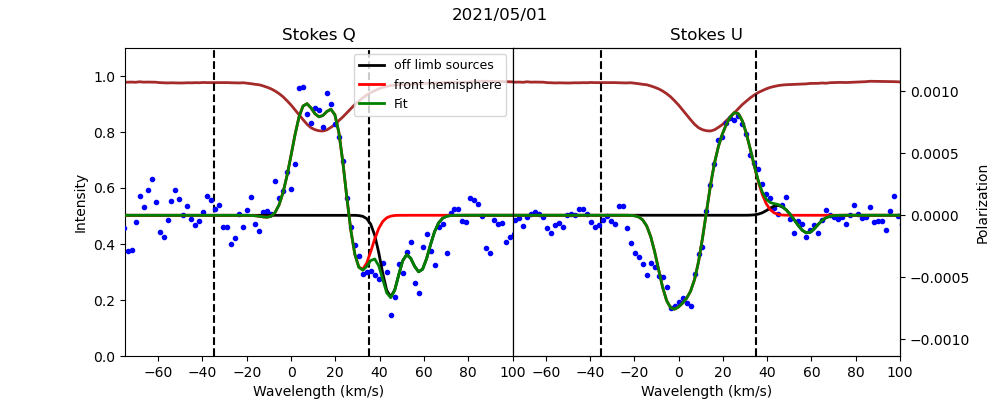}
\caption{Time series of spectropolarimetric observations of $\mu$ Cep corresponding to all dates from September 15, 2020, through May 1, 2021, showing the rise and fall of a convective plume. The meaning of curve colors and styles is the same as in Fig.\ref{prof1}.}
\label{fall}
\end{figure*}

Figure \ref{fall} shows a time series of spectropolarimetric observations of $\mu$ Cep starting on September 15, 2020, and ending on May 1, 2021. The first five observations during the fall and winter of 2020 show the rapid rise of a convective plume above the celestial north limb of the star. This plume can easily be identified in the inferred heights shown on Fig. \ref{height}. Such behavior can also be seen directly in the profiles  as a red peak with negative amplitude in Stokes Q that, day after day, shifts to redder and redder wavelengths, meaning that its projection over the line of sight is greater and greater. Our interpretation of this is that plasma beyond the limb is rising. First, the parts nearer to the limb become visible above the limb and, as time goes on, plasma farther and farther from the limb becomes visible as it reaches the height at which this is geometrically possible. The plume, which is centered well beyond the limb, is rising over a period of 3 months. We lose track of the star from January through April 2021, and in the first observation in May the structure has almost completely disappeared: the polarization beyond the red velocity boundary is small and centered very near the limit, as if only the regions closer to the limb were still emitting light. The plume has disappeared.  We have chosen this event to illustrate how the rise of the plume can be estimated from direct visual inspection of the profiles, before the inversion code confirms the interpretation. The rise of the plume is quite fast, and similarly fast is its disappearance, as  there is barely
any signal of its presence at the opening of the observing window in the following spring. 

It may be tempting to say that the plume fell back into the star, but we have no signature of this. We must recall that, any plasma falling back into the back hemisphere would produce a blueshifted signal that would melt into the signals of rising plasma from the front hemisphere. We have no manner to disentangle to two origins of polarization. 

The event of 2018 is better followed during its disappearance. What we observe is that the signal is still highly visible in the red wing, meaning that the emitted plasma is still rising in the back hemisphere, but its height is lower and lower. We interpret this as follows: the upper parts of the plume of plasma, while still rising, stop emitting light in the atomic lines measured here. This may be because as it cools down, its brightness diminishes, or because atomic lines are no longer excited. Translating our technique to molecular lines, if feasible, would shed light on this. In both cases, the top of the plume cools down first and stops emitting light. We only measure light from the lower parts, which are still hot enough and still raising. This process continues until only the lower parts of the plume are emitting measurable signals. Therefore, at the end of these episodes, we  do not see the plume falling, but just disappearing from our sensing window of atomic lines in what we interpret as a cooling down phase that starts from the top of the plume.

\section{Discussion on the height of the observed structures}

Looking back at Fig. \ref{height} we see that the structure followed in Fig. \ref{fall} reached a minimum height of $1.175R_*$ during those 3 months. This could be higher, because, using geometry, we can only give the lower bound of this height. Taking $1000 - 1200 R_{\odot}$  as the radius of the star, this rise requires an average velocity of 15 to 20 \kms, maintained constant  for 3 months. This velocity fits comfortably with the velocity limit $V_p=70$\kms  in $\mu$ Cep determined for the basic model of the front hemisphere. These are minimum velocities, as we can only determine minimum heights.

The possibility of detecting this kind of plume over the limb, as offered by $\mu$ Cep, is exceptional. The observation of three such events in over five years may suggest that there is some abnormal convective activity in this star, at least when compared with Betelgeuse. Nevertheless, we stick to the assumption that both stars represent different cases of the same physics and that it is just the relatively short span of the observations available that explains the observed differences, and not any fundamental difference between the behaviors of these two RSGs. Building on this assumption,  the measurement of a geometric height made on these structures is deemed typical of  convective plasma features in RSGs, and we generalise it to all other structures imaged on such objects with spectropolarimetry. We consider the  value of $1.1R_*$  as the typical height of the plasma in the atmospheres of RSGs hot enough to emit atomic spectral lines. Making the link with \cite{LA22}, we consider that this measured geometric height, recovered from spectropolarimetry of the deepest atomic lines in the spectrum,  must correspond typically
to the height of their uppermost layer.  Figure 9 of this latter publication must extend therefore up to $1.1R_*$. It is only by considering that this upper layer is visible for several months at that height that in this latter work it is assumed that the observed structures may well have reached $1.3R_*$. 

\section{Conclusion}

Spectropolarimetric observations of the RSG $\mu$ Cep show spectral features in linear polarization that were not observed in the better studied Betelgeuse. Such spectral features are much more redshifted than any other signal and are not permanent features: on certain dates, the observed spectra are qualitatively identical to those of Betelgeuse. We argue that the origin of those unexpected spectral features is convective plumes in the back hemisphere of the star that  rise high enough to be visible above the stellar limb. 

This hypothesis allows us to conserve the inversion algorithms and model that have successfully been applied to Betelgeuse and that have produced images comparable to those from interferometry. However, such a basic model must be extended to allow for temporary sources of polarized light above the stellar limb. We produced an inversion algorithm using this extended model and successfully fitted the observed profiles, including the unexpected new features. This model assumes the presence of a small number of discrete sources over the limb. Although about four such sources are required to correctly fit the profiles, we realize that, at any given date, all those sources appear to be combined to describe a unique but extended source on the star. This observation allowed us to reduce the number of sources over the limb to just two. While the fit with just two sources is not as good as it would with four sources, we still capture the main parameters of the sources and stabilize the convergence of the code, which can be automatically launched to handle the whole dataset available.

The inversion results produce the polar angle position and the height of those sources. This gives us access, for the first time, to a geometric height for the convective structures detected through spectropolarimetry. 

Three events of plasma rising over the limb have been observed during the six years of observation of $\mu$ Cep with Narval and Neo-Narval at the TBL at Pic du Midi. Two of those events were tracked during their rapid rising phase and into their disappearance. The characteristic heights reach  $1.1R_*$ and even $1.175R_*$ in the last observed event. We consider this a typical value of the heights of the convective structures observed in the photosphere of RSGs, and \cite{LA22} use this value to calculate a geometric height from the three-dimensional images of Betelgeuse. Thanks to this measurement, these authors demonstrated that the measured velocities in the plasma are very near the escape velocity of Betelgeuse and that this rising plasma is likely a contributor to the mass loss of these stars.

\begin{acknowledgements}
This work was supported by the "Programme National de Physique Stellaire" (PNPS) of CNRS/INSU co-funded by CEA and CNES.
S.G. acknowledges support under the Erasmus+ EU program for doctoral mobility. S.G. acknowledges partial support by the Bulgarian NSF project DN 18/2.
\end{acknowledgements}

\bibliographystyle{aa}

\bibliography{art72_final}

 \begin{appendix}

 \section{Log of Observations}

 \begin{table} [h]
 \caption{Log of Narval  and Neo-Narval  observations of $\mu$ Cep and polarimetric measurements since July 2015.}
 \begin{tabular}{lcl}

\hline 
\hline
Date &  Julian date & Stokes \\
&&Sequence \\
 \hline
July 10, 2015 & 7214.578 & 8U+8Q \\
September 05, 2015 & 7271.504 & 2U+2Q \\
November 10, 2015 & 7337.462 & 2Q+2U \\
May 16, 2016 & 7525.609 & 2Q+2U \\
June 08, 2016 & 7548.628 & 2U \\
June 16, 2016 & 7556.604 & 2Q \\
June 21, 2016 & 7561.586 & 2U \\
June 27, 2016 & 7567.569 & 2Q \\
July 05, 2016 & 7575.566 & 2Q+2U \\
July 15, 2016 & 7585.517 & 2Q+2U \\
August 02, 2016 & 7603.448 & 2U+2Q \\
September 01, 2016 & 7633.418 & 2Q+2U \\
September 28, 2016 & 7660.436 & 2Q+2U \\
October 07, 2016 & 7669.366 & 4Q+2U \\
November 27, 2016 & 7720.273 & 2Q+2U \\
December 18, 2016 & 7741.265 & 4Q+4U \\
January 07, 2017 & 7761.276 & 2Q+2U \\
April 08, 2017 & 7852.658 & 2Q+2U \\
April 15, 2017 & 7859.664 & 2Q+2U \\
April 22, 2017 & 7866.669 & 2Q+2Q \\
May 07, 2017 & 7881.614 & 2Q+2U \\
May 20, 2017 & 7894.562 & 2Q+2U \\

June 01, 2017 & 7906.57 & 2Q+2U \\

June 11, 2017 & 7916.545 & 2Q+2U \\

June 16, 2017 & 7921.578 & 2Q+2U \\

July 02, 2017 & 7937.627 & 2Q+2U \\

July 11, 2017 & 7946.616 & 2Q+2U \\

July 31, 2017 & 7966.561 & 2Q+2U \\

August 08, 2017 & 7974.536 & 2Q+2U \\

August 13, 2017 & 7979.45 & 2Q+2U \\

August 21, 2017 & 7987.524 & 2Q+2U \\

September 02, 2017 & 7999.445 & 2Q+2U \\

September 05, 2017 & 8002.412 & 2Q+2U \\

September 13, 2017 & 8010.449 & 2Q+2U \\

September 20, 2017 & 8017.504 & 2Q+2U \\

September 27, 2017 & 8024.37 & 2Q+2U \\

October 02, 2017 & 8029.469 & 2Q+2U \\

October 07, 2017 & 8034.342 & 2Q+2U \\

October 12, 2017 & 8039.359 & 2Q+2U \\

October 30, 2017 & 8057.341 & 2Q+2U \\

November 07, 2017 & 8065.285 & 2Q+2U \\

November 14, 2017 & 8072.337 & 2Q+2U \\

November 19, 2017 & 8077.255 & 2Q+2U \\

November 26, 2017 & 8084.269 & 2Q+2U \\

December 04, 2017 & 8092.275 & 2Q+2U \\

May 17, 2018 & 8256.624 & 2Q+2U \\

June 14, 2018 & 8284.607 & 2Q+2U \\

June 30, 2018 & 8300.503 & 2Q+2U \\

July 22, 2018 & 8322.619 & 2Q+2U \\
\hline
\end{tabular}
\end{table}
 \begin{table} [t]
\begin{tabular}{lcl}

\hline 
\hline
Date &  Julian date & Stokes \\
&&Sequence \\
 \hline

August 13, 2018 & 8344.677 & 2Q+2U \\

September 26, 2018 & 8388.478 & 2Q+2U \\

October 25, 2018 & 8417.442 & 2Q+2U \\

November 14, 2018 & 8437.322 & 2Q+2U \\

December 10, 2018 & 8463.313 & 2Q+2U \\

January 04, 2019 & 8488.239 & 2Q+2U \\

January 16, 2019 & 8500.281 & 2Q+2U \\

March 21, 2019 & 8564.685 & 2Q+2U \\

May 05, 2019 & 8609.594 & 2Q+2U \\

June 01, 2019 & 8636.633 & 2Q+2U \\

June 18, 2019 & 8653.622 & 2Q+2U \\

July 18, 2019 & 8683.555 & 2Q+2U \\

August 02, 2019 & 8698.589 & 2Q+2U \\

August 15, 2019 & 8711.498 & 2Q+2U \\

August 30, 2019 & 8726.552 & 2Q+2U \\

January 06, 2020 & 8855.288 & 2Q+2U \\

May 17, 2020 & 8987.572 & 1Q+2U \\

June 22, 2020 & 9023.621 & 2Q+2U \\

July 05, 2020 & 9036.583 & 2Q+2U \\

July 24, 2020 & 9055.576 & 2Q+2U \\

August 22, 2020 & 9084.518 & 2Q+2U \\

September 15, 2020 & 9108.483 & 2Q+2U \\

October 16, 2020 & 9139.313 & 2U+2Q \\

November 21, 2020 & 9175.329 & 2Q+2U \\

December 18, 2020 & 9202.256 & 2Q+2U \\

January 13, 2021 & 9228.271 & 2Q+2U \\

May 01, 2021 & 9337.647 & 2Q+2U \\

May 26, 2021 & 9361.603 & 2Q+2U \\

June 14, 2021 & 9380.536 & 2Q+2U \\

July 10, 2021 & 9406.619 & 2Q+2U \\

August 07, 2021 & 9434.455 & 2Q+2U \\

August 19, 2021 & 9446.59 & 2Q+2U \\

September 04, 2021 & 9462.43 & 2Q+2U \\

October 06, 2021 & 9494.441 & 2Q+2U \\

October 13, 2021 & 9501.405 & 2Q+2U \\

November 09, 2021 & 9528.363 & 2Q+2U \\

December 13, 2021 & 9562.279 & 2Q+2U \\

December 22, 2021 & 9571.252 & 2Q+2U \\

January 11, 2022 & 9591.264 & 2Q+2U \\
\hline
\end{tabular}
\label{tab1}

\textbf{Notes:} Columns give the date, the heliocentric Julian date (+2\,450\,000),  and the observed Stokes sequence, that is, how many observations of which Stokes parameter were made at that date. An observation consists of four exposures with changing polarimetric modulation that, after reduction, produce polarization spectra of either Stokes $Q$ or $U$. Beyond a 2 year proprietary embargo, all data are publicly available at PolarBase (http://polarbase.irap.omp.eu/).

\end{table}

 \end{appendix}

\end{document}